\documentclass[doublecol,linenumbers]{epl2}
\usepackage{graphicx}
\usepackage[english]{babel}
\usepackage{latexsym,amsmath,amscd,amssymb,graphicx,amsbsy,color,xcolor,amsfonts,amsthm}

\title{Predicting the relativistic periastron advance of a binary without curving spacetime.}
\shorttitle{Predicting { binary periastron advance}}
\author{ Y. Friedman, S. Livshitz and J. M. Steiner}
\institute{
   Jerusalem College of Technology\\Jerusalem, Israel}

\abstract{
 Relativistic Newtonian Dynamics, the simple  model used previously for predicting accurately the anomalous precession of Mercury, is now applied to predict the  periastron advance of { a binary}. The classical treatment of {a}  binary as a two-body problem is modified to account for the influence of the gravitational potential on spacetime. {Without curving spacetime, the model predicts the identical equation for the relativistic periastron advance as the post-Newtonian approximation of general relativity formalism} thereby providing {further} substantiation of this model.}

\begin{document}
\maketitle

\section{Introduction}

The predictions of the anomalous precession of Mercury, the periastron advance of the Hulse-Taylor binary and{ of the most relativistic double pulsar PSR J0737-3039A/B} are considered amongst the major proofs of  Eintein's theory of General Relativity (GR). In this paper we apply the simple relativistic model termed as Relativistic Newtonian Dynamics (RND) to predict the periastron advance of {any} binary and its origin. This model incorporates the influence of the gravitational potential on spacetime in Newtonian gravity without the need of curving the spacetime. The model was previously explored in \cite{F16} and applied \cite{FS} to predict accurately the anomalous precession of Mercury.

These two predictions indicate that RND can provide an alternative to GR for problems involving  gravitation. Since in a gravitational field different objects positioned at the same point in  spacetime follow the same trajectory, the RND trajectories can be viewed as geodesics in a curved  spacetime, as in GR (geometric theory of gravitation). As it was shown in \cite{FS} for planetary motion the RND trajectories are the geodesics of the Schwarzschild metric. In non gravitational fields, however, where different objects positioned at the same point in spacetime follow different trajectories, such geometric model no longer applies. The RND model, nevertheless, is also applicable to these non-gravitational fields.

For gravitational fields it is known  \cite{MTW} that \begin{enumerate}
                                                        \item the gravitational redshift (time dilation due to gravitational potential), can be derived solely from energy conservation and Planck's equation,
                                                        \item the existence of the gravitational redshift shows that a consistent theory of gravity  cannot be constructed within the framework of SR.
                                                      \end{enumerate}
For non-gravitational fields \begin{enumerate}
                                    \item the  energy conservation and Planck's equation also predict a time shift depending on the position in space,
                                    \item such time dilation cannot be described within the framework of SR.
                                  \end{enumerate}
Indeed, SR defines time dilation due to velocity, but not due to position.  We believe that, for both the above fields, SR does not explain the \textit{position} dependent time dilation, for the reason that it does not consider the influence of the potential energy on spacetime.  This is is the relativistic basis of  RND.

\section{Classical two-body problem}

Consider two objects {$S_1,S_2$} with masses $m_1$ and $m_2$  positioned in an inertial system with position vectors $\mathbf{r}_1$ and $\mathbf{r}_2$ respectively with a force along the line joining them. Denote by $\mathbf{r}=\mathbf{r}_2-\mathbf{r}_1$ the displacement vector between them and by $\hat{\mathbf{r}}$ the unit vector in the direction $\mathbf{r}$. Let $\mathbf{F}=F\hat{\mathbf{r}}$ be the force acting on the first object. By Newton's third law the force acting on the second object is $-\mathbf{F}.$
The acceleration of these objects by Newton's second law  are respectively
\begin{equation}\label{two_body_dyn}
\ddot{\mathbf{r}}_1 = \frac{1}{m_1}\mathbf{F},\;\;\;
\ddot{\mathbf{r}}_2 =-\frac{1}{ m_2}\mathbf{F}.
                                        \end{equation}

Note that the reciprocal of the mass, the constant of proportionality relating the force to the acceleration it causes, expresses the ``agility" of the object to be accelerated.
From (\ref{two_body_dyn}), the relative acceleration of $\mathbf{r}$ is
\begin{equation}\label{r_ddot}
  \ddot{\mathbf{r}}=-\left(\frac{1}{m_1}+\frac{1}{m_2}\right)\mathbf{F}.
\end{equation}
Since the accelerations of the two objects are in opposite directions, the magnitude of their relative acceleration $\ddot{\mathbf{r}}$ is the sum of acceleration magnitudes of each object. Furthermore, since the forces generating these accelerations have the same magnitude, the agility of the object pair is the sum of their respective agility, as seen from
 (\ref{r_ddot}).

  This agility of the object pair is the agility of a single fictitious object $P'$ with reduced mass
   \begin{equation}\label{rho}
  \rho=\frac{m_1m_2}{m_1+m_2} .
\end{equation}
 Thus, the evolution of the relative position $\mathbf{r}$ in the two body problem can obtained from the dynamics of $P'$ under the force $\mathbf{F}$.

 From equations (\ref{two_body_dyn}),  $ m_1\ddot{\mathbf{r}}_1+ m_2\ddot{\mathbf{r}}_2=0$ implying that the center of mass defined by
 \begin{equation}\label{centerM}
  \mathbf{R}=\frac{m_1{\mathbf{r}}_1+ m_2{\mathbf{r}}_2}{m_1+m_2}
\end{equation}
moves uniformly in  a straight line. We define a new inertial frame by $K$ by translating the origin of the original frame to $O=\mathbf{R}$. Thus, the relative position of each object in $K$ is given by
\begin{equation}\label{Two_traj}
 \tilde{\mathbf{r}}_1=\mathbf{r}_1-\mathbf{R}=\frac{-m_2}{m_1+m_2}\mathbf{r},\;\;
  \tilde{\mathbf{r}}_2=\mathbf{r}_2-\mathbf{R}=\frac{m_1}{m_1+m_2}\mathbf{r}.
\end{equation}

%\section{Classical treatment of the Hulse-Taylor binary as a two-body problem}

The gravitational force between {two objects in a binary with}  masses $m_1, m_2$, respectively,  is given by $\mathbf{F}=-\frac{Gm_1m_2}{r^2}\hat{\mathbf{r}}$, where $G$ is the gravitational constant.  Equation (\ref{r_ddot}) becomes
\begin{equation}\label{r_ddot_grav}
  \rho \ddot{\mathbf{r}}=-\frac{GM\rho}{r^2}\hat{\mathbf{r}},\;\;\; M=m_1+m_2,
\end{equation}
which is the equation of motion of a fictitious planet $P'$ of mass $\rho$ in the central gravitational force field of a massive fictitious ``Sun" $S$  with mass $M$ at the  origin $O$ of $K$ and relative position $\mathbf{r}$.

Using that the potential energy of the gravitational field is $U=-\frac{GM\rho}{r}$, the energy conservation equation is
\begin{equation}\label{Energy}
\frac{\rho}{2}\left(\frac{d\mathbf{r}}{dt}\right)^2-\frac{GM\rho}{r}=E
\end{equation}
expressing that the total energy $E$ (the sum of the kinetic and potential energies) of $P'$ is conserved on its orbit.

 Note that from (\ref{rho}) and (\ref{Two_traj}), the kinetic energy of $P'$
\begin{equation*}
  \frac{\rho}{2}\left(\frac{d\mathbf{r}}{dt}\right)^2=\frac{m_1}{2}\left(\frac{d\tilde{\mathbf{r}}_1}{dt}\right)^2+\frac{m_2}{2}\left(\frac{d\tilde{\mathbf{r}}_2}{dt}\right)^2
\end{equation*}
is the sum of the kinetic energies of the pulsar and its companion in $K$ and the potential energy of $P'$
\begin{equation}\label{PotEng}
  U(\mathbf{r})=-\frac{GM\rho}{r}=-\frac{Gm_1m_2}{r}
\end{equation}
 defines properly the force $\mathbf{F}$.

Dividing  equation (\ref{Energy})  by  $\frac{\rho c^2}{2},$  where $c$ is the speed of light we obtain the dimensionless  energy conservation equation
\begin{equation}\label{RE decompNG0}
  \frac{1}{c^2}\left(\frac{d\mathbf{r}}{dt}\right)^2-\frac{r_s}{r}=\mathcal{E},
\end{equation}
where
\begin{equation}\label{Shuart}
 r_s=\frac{2GM}{c^2}
\end{equation}
is the Schwarzschild radius of $S$ which is also the Schwarzschild radius of the binary (the minimal distance between the {objects for which}) the relative velocity to separate them is less than $c$). The  \textit{dimensionless kinetic energy}  is $\frac{1}{c^2}\left(\frac{d\mathbf{r}}{dt}\right)^2=\beta^2$, where $\beta$ is the known beta-factor and
 the absolute value of the \textit{dimensionless potential energy}  is
\begin{equation}\label{u_def}
  u=\frac{2GM}{rc^2}=\frac{r_s}{r}.
\end{equation}
Finally, we denote by $\mathcal{E}=\frac{2E}{\rho c^2}$ the \textit{dimensionless total energy} of the orbit.

Using (\ref{Two_traj})
 \begin{equation}\label{ang_moment_binary}
   \rho \mathbf{r}\times \dot{\mathbf{r}}=m_1\tilde{\mathbf{r}}_1\times \dot{\tilde{\mathbf{r}}}_1+m_2\tilde{\mathbf{r}}_2\times \dot{\tilde{\mathbf{r}}}_2,
\end{equation}
showing that in $K$ the angular momentum of $P'$ and that of the binary with respect to $O$ are the same. From (\ref{r_ddot_grav}) follows that the angular momentum per unit mass $\mathbf{J}$ of $P'$ is conserved on the orbit implying that $\mathbf{r}$ is in the plane perpendicular to $\mathbf{J}$ and from (\ref{Two_traj}) the {trajectories of the two objects} are in this plane.

We introduce polar coordinates $r,\varphi$ in this plane with origin $O$, where $\varphi$ is the dimensionless polar angle, measured in radians. Conservation of angular momentum per unit mass $J$, allows us to express the angular velocity as
\begin{equation}\label{Ang_vel}
  \frac{d\varphi}{dt}=\frac{J}{r^2}
\end{equation}
 and to decompose the square of the  velocity of the planet as the sum of the squares of its  orthogonal \textit{radial} and \textit{transverse} components
 \begin{equation}\label{Theta1_dot}
 \left(\frac{d\mathbf{r}}{dt}\right)^2=\left(\frac{dr}{dt}\right)^2+\frac{J^2}{r^2}.
\end{equation}
 Substituting this into (\ref{RE decompNG0}) we obtain the  classical \textit{dimensionless energy conservation equation}
 \begin{equation}\label{RE decompNG}
 \frac{1}{c^2}\left(\frac{dr}{dt}\right)^2 +\frac{J^2}{c^2r^2}-\frac{r_s}{r}=\mathcal{E}.
\end{equation}

Using the definition (\ref{u_def}) of $u$, and denoting its derivative with respect to $\varphi$ by $u'$, it can be shown that
\begin{equation}\label{uprime_vel}
\frac{dr}{dt}=-\frac{J}{r_s} u'. \end{equation}
Hence, equation (\ref{RE decompNG}) becomes
\begin{equation}\label{ClassEnergy}
\frac{J^2}{r_s^2 c^2}\left((u')^2+ u^2\right)= u +\mathcal{E}.
\end{equation}
 Multiplying this equation by $2\mu$, where $\mu$ is a\textit{ unit-free orbit parameter}
\begin{equation}\label{zeta}
\mu=\frac{r_s^2 c^2}{2J^2}
\end{equation}
 we obtain
\begin{equation}\label{u_prime}
 (u')^2=-u^2+2\mu u +2\mu \mathcal{E}.
\end{equation}

Differentiating this equation with respect to $\varphi$ and dividing by $2u'$ we obtain a linear differential equation with constant coefficients
\begin{equation}\label{ClasFinal}
  u''+u=\mu.
\end{equation}
Its solution is
\begin{equation}\label{SolU}
  u(\varphi)=\mu(1+\varepsilon\cos (\varphi-\varphi_0)),
\end{equation}
where $\varepsilon$ - the eccentricity of the orbit, and $\varphi_0$ - the polar angle of the perihelion. This implies that
\begin{equation}\label{ClasOrbit}
  r(\varphi)=\frac{r_s/\mu}{1+\varepsilon \cos (\varphi-\varphi_0)}
\end{equation}
and the orbit is a \textit{non-precessing} ellipse. Since the minima of $r(\varphi)$, corresponding to the perihelion, occur when $\varphi=\varphi_0+2\pi n$, $n=0,1,2,\cdots$ the position of the perihelion will not change with the revolution of the planet $P'$.

 The orbit constant $\mu$, as we shall see later, plays a major role for precession of the orbit. It has both a physical and geometric meaning. From equation (\ref{SolU})
\begin{equation}\label{mu_meaning}
  \mu=\frac{1}{2\pi}\int_0^{2\pi}u(\varphi)d\varphi,
\end{equation}
the absolute value of the angular average  dimensionless potential energy of $P'$ on the orbit. Moreover, from (\ref{ClasOrbit}) $\mu = \frac{r_s}{L},$ where $L$ is the semi-latus rectum of the orbit of $P'$. From (\ref{u_def}), the $u$ values on the orbit achieve its maximum and minimum values $u_p$  and a $u_a$ at the perihelion and aphelion, respectively. Thus,  from (\ref{SolU})
\begin{equation}\label{elipse_par}
\mu=\frac{u_p+u_a}{2}.
\end{equation}

Hence, from (\ref{Two_traj}),  the orbits of the {two objects} are non-precessing ellipses, see Figure 1. The apastron separation of the binary (the maximum distance between the {the two objects}) is the same as the apogee (maximal distance of $P'$ from $O$). Similarly, the periastron separation (the minimum distance between the {two objects}) is the same as the perigee (minimal distance of $P'$ from $O$).

\begin{figure}[h!]
\centering
 \scalebox{0.6}{\includegraphics{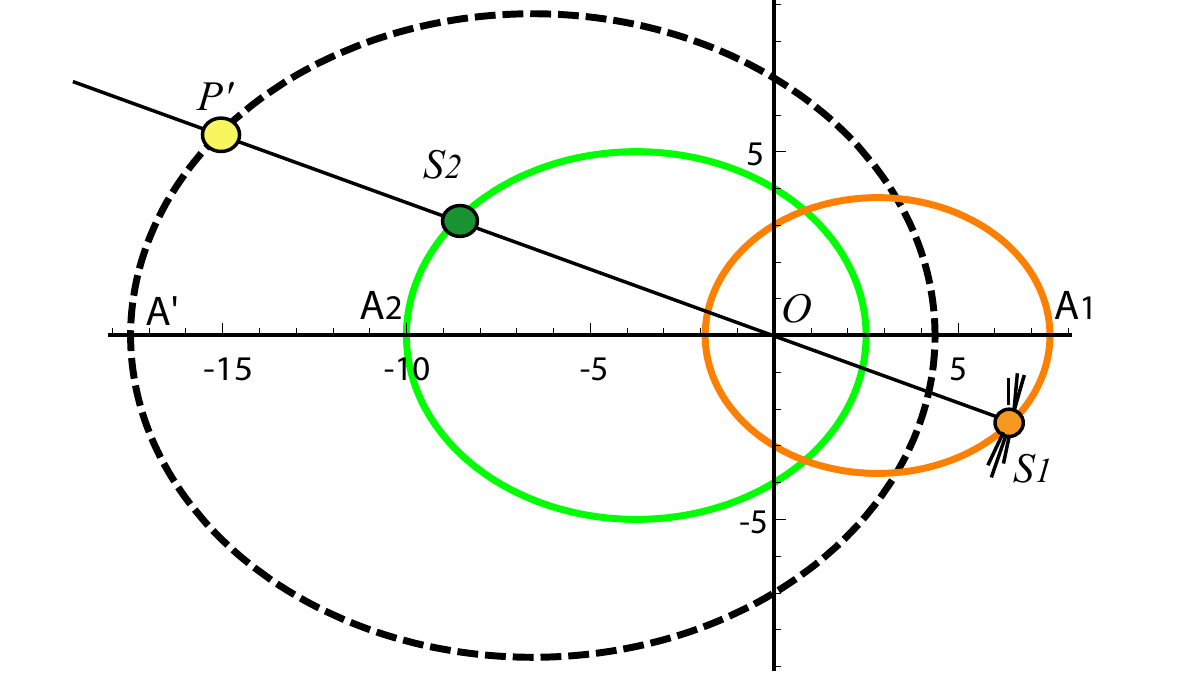}}
 \caption{The orbits of the {the two objects} $S_1,S_2$ (in red and green) and the trajectory of the  fictitious planet  $P'$ (in black dashed). The apastron separation $|A_1,A_2|$ of the binary equals to apogee $A'$ of $P'$. Point $O$ is the center on mass $\mathbf{R}$ for the binary and position of $S$. }\label{2 states}
\end{figure}

%
%The masses of the pulsar and its companion in the Hulse-Taylor binary  are $m_1=2.8676\cdot 10^{30}$ [kg] and $m_2=2.7661\cdot 10^{30}$kg. Thus, the Schwarzschild radius of the pair is
%\begin{equation*}
% r_s =\frac{2GM}{c^2} =\frac{2\cdot 6.674\cdot 10^{-11}\cdot 5.6337\cdot10^{30}}{9\cdot10^{16}}=8,355.4 m,
%\end{equation*}
%the binaries apastron and perriastron are $r_a=3.153\cdot10^{9}m$ and $r_p=0.746 \cdot10^{9}m$, respectively, implying that
% \begin{eqnarray*}
%  \nonumber
%  u_a &=&\frac{r_s}{r_a}= \frac{8,355.4}{3.153 \cdot10^{9}}=2.6500\cdot10^{-6}\\
%u_p &=& \frac{r_s}{r_p}= \frac{8,355.4}{0.746 \cdot10^{9}}=1.1200\cdot10^{-5}.
% \end{eqnarray*}
%From this and  (\ref{elipse_par}) we get
%\begin{equation*}
%  \mu=\frac{1.1200\cdot10^{-5}+2.6500\cdot10^{-6}}{2}=6.925\cdot10^{-6}.
%\end{equation*}

\section{Potential energy as the source of precession}

The classical Newtonian solution for a binary does not account for the periastron advance. In fact, the observed Hulse-Taylor pulsar's periastron advance in one day is approximately as the one observed previously for Mercury in 100 years. { Nowadays, there are ten neutron star binaries for which the advance of periastron has been measured.  Among all those, the double pulsar PSR J0737-3039A/B is the most relativistic with a periastron advance of 16.899 degrees/year.}  We will show that {the} periastron advance is a result of the influence of the gravitational potential (\ref{PotEng}) on the spacetime.

  Rather then proposing an a priori postulate for such influence, we will be guided by the Einstein's Equivalence Principle and Clock Hypothesis.
 Einstein's Equivalence Principle states \cite{Einstein07} the complete physical equivalence of a gravitational field and a corresponding accelerated system. To be able to use this principle effectively, we use the notion of escape velocity and escape trajectory. For any  displacement $\mathbf{r}$ of a binary the escape velocity $\mathbf{v}_e(\mathbf{r})$ is the minimal relative velocity needed to  separate the binary. From symmetry consideration $\mathbf{v}_e(\mathbf{r})$ is in the direction of $\mathbf{r}$. Define the \textit{escape trajectory} as an imaginary (de)accelerated trajectory of a test object starting at point $P_0$ positioned at $\mathbf{r}$  with escape velocity $\mathbf{v}_e(\mathbf{r})$, and progressing freely in a  decreasing potential field $U(r)$ to the ultimate point $P_\infty$ with zero potential. Note that on this trajectory we have equality of the kinetic and potential energies.

Using the Equivalence Principle the effect of gravitation on spacetime is modeled by the (de)accelerating system attached to escaping test object. Denote by $K_0, K_\infty$ the comoving spacetime reference frames at $P_0$ and $P_\infty$, respectively. Note that $K_\infty$ is an inertial (lab) frame resting in $K$, hence with the same spacetime. Using an extension of Einstein's Clock Hypothesis, (extending the time to spacetime transformation) introduced in  \cite{Mash1,Mash2},  we can interconnect the spacetime transformations between the two accelerated frames at $P_0$ and $P_\infty$ by use of the Lorentz transformation from $K_0$ to $K_\infty$.

 For a central force potential, the frame $K_0$  moves with escape velocity $\mathbf{v}_e $ in the radial direction with respect to $K$. We choose the first spacial coordinate in the radial direction, and denote by $\beta_e^2=v_e^2/c^2=u$, where $u$ is defined by (\ref{u_def}). Then,  the spacetime transformation  (Lorentz transformation) from $K_0$ to $K_\infty$ is
 \begin{eqnarray}\label{FFFbasis}
    \nonumber ct &=& \tilde{\gamma}(ct'+\beta_ex'_1),\;\;\; x_2 = x'_2,\\
    x_1 &=& \tilde{\gamma}(\beta_e t'+x'_1),\;\;\; \;x_3 = x'_3,
    \end{eqnarray}
 where
 \begin{equation}\label{gammatilde_def}
 \tilde{\gamma}=\frac{1}{\sqrt{1-v_e^2/c^2}}=\frac{1}{\sqrt{1-u}}
\end{equation}
 is the known gravitational \textit{time dilation factor}.

 From the transformation formulas (\ref{FFFbasis}) we get
 \begin{equation} \Delta t=\tilde{\gamma}\Delta t',\;\Delta x_1=\tilde{\gamma}\Delta x'_1,\;\Delta x_2=\Delta x'_2,\;\Delta x_3=\Delta x'_3\,\end{equation}
 implying that the 3D velocity transformations between these systems is
 \begin{equation}\label{valoc_transf}
   (v_1,v_2,v_3) =(v'_1,\tilde{\gamma}^{-1}v'_2,\tilde{\gamma}^{-1}v'_3).
 \end{equation}
Thus the influence of acceleration or potential energy at $x_0$ on any velocity is expressed by multiplication of the component of this velocity transverse to  $\nabla U(P_0)$ by $\tilde{\gamma}^{-1},$ where the time dilation factor $\tilde{\gamma}$ is defined by (\ref{gammatilde_def}).

 \section{Relativistic Newtonian Dynamics {application to binaries} }

\textit{Relativistic Newtonian Dynamics} (RND) is a modification of the Newtonian dynamics by transforming it from absolute space and time to spacetime influenced by energy.
For a central force, the direction influenced by the potential energy is the radial direction implying that the radial velocities are not affected by this influence, while the transverse ones should be  multiplied by $\tilde{\gamma}^{-1}=\sqrt{1-u(r)}$. This implies, that in our lab frame $K$ the decomposition of the square of the  velocity of the particle as the sum of the squares of its  orthogonal radial and transverse components, given by (\ref{Theta1_dot}), should be modified to
\begin{equation}
 \left(\frac{d\mathbf{r}}{dt}\right)^2=\left(\frac{dr}{dt}\right)^2+\frac{J^2}{r^2}(1-u(r)).
\end{equation}

Our model also reveals the source of the precession of the planetary orbit. As mentioned above, in NG, the radial and the transverse periods are identical, resulting in a non-precessing orbit. In SR, \textit{both} the radial and transverse components of the velocity are altered, resulting in unequal periods with relatively small difference between them and hence a small precession. In our  model, \textit{only} the radial component of the velocity is influenced, while the transverse (angular) component is not. This, in turn,  accentuates the difference between these periods, resulting in the  observed precession, as follows.
Thus, considering the influence of the potential energy, the dimensionless energy conservation equation (\ref{RE decompNG}) becomes
\begin{equation}\label{RND decompNG}
 \frac{1}{c^2}\left(\frac{dr}{dt}\right)^2 +\frac{J^2}{c^2r^2}(1-u(r))-u(r)=\mathcal{E}.
\end{equation}
This equation together with equation (\ref{Ang_vel}) form a first order system of differential equations with respect to $r(t),\varphi(t)$. They are the \textit{RND equations of motion under a central force}.

For an inverse-square law force, using known methods (see for example \cite{KEK}, \cite{GS} and \cite{FS}) we can obtain the trajectory of the motion by solving equation (\ref{RND decompNG}) for the function $u(\varphi)$. If we denote $u'=\frac{du}{d\varphi}$, then substituting (\ref{uprime_vel}) in the above equation gives
\begin{equation}\frac{J^2}{c^2r_s^2}(u')^2 +\frac{J^2u^2}{c^2r_s^2}(1-u)-u=\mathcal{E}.\end{equation}
Multiplying this equation by $2\mu$, where $\mu$ is defined by (\ref{zeta}),
we obtain
\begin{equation}\label{uprime_eqn}
  (u')^2 = u^3-u^2 +2\mu u +2\mu \mathcal{E}.
\end{equation}
This equation is identical to (\ref{u_prime}) in {NG, except that it has a} very small (since $u\ll 1$) additional term $u^3$ on the right-hand side. {This result is the same result as that of GR.}
\begin{figure}[h!]
\centering
 \scalebox{0.6}{\includegraphics{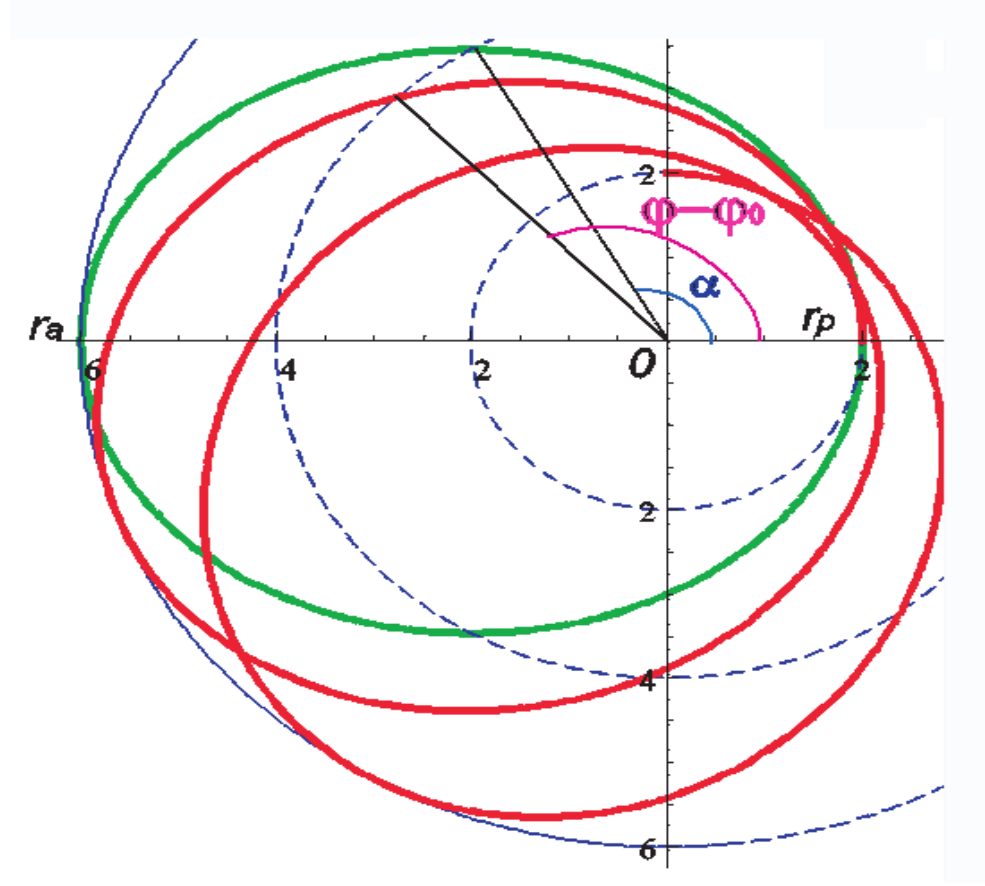}}
 \caption{The precessing orbit  of the {fictitious planet $P'$} in red. The classical orbit in green. }\label{Prec}
\end{figure}

We seek a solution of this equation in the form generalizing (\ref{SolU}),
\begin{equation}\label{GRu_form}
  u(\varphi)=\mu(1+\varepsilon\cos\alpha(\varphi))
\end{equation}
for some function $\alpha(\varphi)$. As before, two roots of the cubic on the right-hand side of (\ref{uprime_eqn}), are the $u$ values  $u_p$ and $u_a$ of the the perihelion and aphelion, respectively.  Moreover, since the coefficients of this cubic are constant for a given orbit, these values will not change from one revolution to the next. We denote the third root of this cubic by $u_e$. Thus, equation (\ref{uprime_eqn}) can be factorized as
\begin{equation}\label{u_primeGRprod}
 ( u')^2=-(u-u_p)(u-u_a)(u_e-u).
\end{equation}

 From equation (\ref{GRu_form}), $( u')^2=(\alpha')^2\mu^2\varepsilon^2\sin^2\alpha(\phi)$, $u_p=\mu+\mu\varepsilon$ and $u_a=\mu-\mu\varepsilon.$
 Moreover, since the sum of the roots of this cubic is 1, \begin{equation}u_e=1-(u_p+u_a)=1-2\mu.\end{equation}
Substituting  these into (\ref{u_primeGRprod}), yields after simplification
 \begin{equation}\alpha'=\frac{d\alpha}{d\varphi} =(1-3\mu-\mu\varepsilon\cos\alpha(\varphi))^{1/2}.\notag\end{equation}
 This allows us to obtain the dependence of $\varphi$ on $\alpha$ as
 \begin{equation}\label{phy_alpha}
   \varphi(\alpha) =\varphi_0+\int_0^\alpha (1-3\mu-\mu\varepsilon\cos\tilde{\alpha})^{-1/2}d\tilde{\alpha}.
 \end{equation}

As we have shown in \cite{FS} and \cite{F16}, {this yields  the precession $3\pi \mu \frac{rad}{rev}$ of $P'$.  As mentioned above, this  is also the relativistic  periastron advance per revolution of the  binary, the object we are looking for. Hence, the commonly used non-Keplerian  parameter $\dot{\omega}$-the periastron advance per unit of time or time rate of change (precession) of the longitude of the periastron, is
\begin{equation}\label{prec1}
 \dot{\omega}=3\pi \frac{\mu }{P_b},
\end{equation}
where $P_b$ is the orbital period of the binary per same unit of time.}

{ We express this formula for in terms of the Keplerian parameters of the orbits of the two objects of the binary. From (\ref{u_def}) and (\ref{elipse_par})
\begin{equation*}
  u_p=\frac{2GM}{r_pc^2}, \; u_a=\frac{2GM}{r_ac^2}\;\Rightarrow\;\mu=\frac{GM}{c^2}\left(\frac{1}{r_p}+\frac{1}{r_a}\right),
\end{equation*}
where $r_p,r_a$ denote the periastron and apastron separations, respectively.

As evident from  (\ref{Two_traj},) the eccentricity of $P'$ is the same as that of the eccentricity of each object. Hence, we can express $r_p$ and $r_a$ in terms of $a_1,a_2$ are the semimajor axes of objects 1 and 2 respectively, as
\begin{equation*}
 r_p= (1-\varepsilon)(a_1+a_2),\;\;r_a= (1+\varepsilon)(a_1+a_2),
\end{equation*}
implying that
\begin{equation}\label{1/r}
  \frac{1}{r_p}+\frac{1}{r_a}=\frac{2}{a}(1-\varepsilon^2)^{-1},
\end{equation}
 where $a=a_1+a_2$ is the semimajor axis of $P'$.
From Kepler's formula for $P'$,
 \begin{equation*}
  P_b=2\pi\sqrt{\frac{a^3}{GM}},\;\; \Rightarrow\;\;a=(GM)^{1/3} \left(\frac{P_b}{2\pi}\right)^{2/3}
\end{equation*}
Substituting all these into (\ref{prec1})
\begin{equation*}
 \dot{\omega}=3\pi\frac{2GM}{c^2 P_b}(GM)^{-1/3} \left(\frac{P_b}{2\pi}\right)^{-2/3}(1-\varepsilon^2)^{-1}.
\end{equation*}
and
\begin{equation}\label{Final}
 \dot{\omega} =3\frac{(GM)^{2/3}}{c^2(1-\varepsilon^2)}\left(\frac{P_b}{2\pi}\right)^{-5/3},
\end{equation}
which is the post-Keplerian equation for the relativistic advance of the periastron $\dot{\omega}$, given for example in \cite{Kramer}, \cite{Lorimer}, \cite{Weisb} and{ \cite{DD}}.

This formula can be used to provide the total mass $M$ of the system, which combined with the theory independent mass ratio, yields the individual masses of the system.}
 %Thus, the  Hulse-Taylor pulsar's periastron advance per revolution is
% \begin{equation*}
%  d\varphi= 3\cdot180\cdot 6.925\cdot10^{-6} =0.0374^\circ .
%\end{equation*}
% Using that the period of the binary is $T= 7.7519$ hours this give the precession per year
% $d\varphi\cdot 365\times24/T=4.226^\circ$ per year, which is the observed one.

 \section{Summary and Discussion}

 We have {derived the  periastron advance formula (\ref{Final}) of any binary} by the use of our Relativistic Newtonian Dynamics (RND) which incorporates the influence of the potential energy on flat spacetime. { This formula is the same as the one obtained using the first order Post-Newtonian (PN) approximation (1PN) of GR.}

{ The PN approximation is effectively an expansion of Einstein's theory in powers of a small parameter $\dot{r}^2/c^2\sim GM/(c^2 r)$. The periastron advance of a binary using the 2PN approximation was obtained in \cite{DS} by employing a second PN reduced Hamilton function $\hat{H}$ in isotropic coordinates in the center of mass system. The 3PN approximation for this advance  was derived in \cite{KG}.

The difference between the RND and the PN approach is best seen by comparing our RND reduced Hamilton function $H_{rn}$ in  flat space in the center of mass system, with the corresponding $\hat{H}$. As evident from (\ref{RND decompNG}) and (\ref{u_def}),
\begin{equation}\label{ERD_Ham}
  H_{rn}(\mathbf{r},\dot{\mathbf{r}},M)=\frac{\dot{\mathbf{r}}^2}{2}-\frac{GM}{r}-\frac{GM}{rc^2}\left(\dot{\mathbf{r}}^2-(\dot{\mathbf{r}}\cdot \mathbf{n})^2\right),
\end{equation}
where $\mathbf{n}$ is a unit vector in the radial direction.

Unlike $\hat{H}$, our $H_{rn}$ is derived from the principles of our model and is not an approximation. The relativistic contribution in $H_{rn}$ embodied in the last term of (\ref{ERD_Ham})  resembles  similar 1PN terms of $\hat{H}$, however $\hat{H}$ involves two extra 1PN terms with complicated coefficients. The difference in the complexity between $\hat{H}$ and $H_{rn}$ results partly by the use of isotropic coordinates instead of flat spacetime, and partly by the parametrization of the trajectory by proper time instead of time. Since the parametrization of the trajectory does not affect the formula for the periastron advance, one may use any parametrization which may simplify the solution.

As it was shown in  \cite{Bagchi}, for more relativistic binaries such as a neutron star with a 10 solar mass black hole companion, the 2PN approximation and the leading order spin-orbit coupling term could have significant contributions to the periastron advance of such binaries. Actual observations of such binaries, once  available,  will provide the real test for the accuracy of RND.}

 It is known  \cite{MTW} that the gravitational redshift, can be derived from energy conservation (and using Planck's equation) and that "an argument of A. Schild yields an important conclusion: the existence the gravitational redshift shows that a consistent theory of gravity  cannot be constructed within the framework of special relativity". This is because special relativity does not consider the influence of potential energy on spacetime.  RND predicts accurately both the anomalous precession precession of Mercury \cite{FS} as well as the periastron advance {of any binary}. This indicates that RND provides an alternative to General Relativity (GR) for problems involving  gravitation.

 Note that in both problems, we considered, there exits a preferred reference frame: for a central force this is the frame attached to the center of the force, while for a binary this is the frame attached to the center of mass of the binary. Thus, in order to simplify the solution, we can work in these frames and use space and time separately, without using the 4D formulation. Only for deriving the influence of the gravitational potential on spacetime we have to use the 4D formulation.

 Since in gravity different objects positioned at the same spacetime follow the same trajectory, the RND dynamics, which considers the change of the spacetime at any point in space, can be expressed as curving of spacetime, as in GR (geometric theory of gravitation). As it was shown in \cite{FS} for planetary motion the RND trajectory is the same as the geodesic under a Schwarzschild metric.

 For non gravitational potentials for which different objects positioned at the same spacetime follow different trajectories, the geometric model does not express the influence of these potentials on spacetime. As stated in Living Review \cite{CW}, if one were to use Einstein's  GR for non-gravitational fields, it is currently assumed that the non-gravitational laws of physics are written in the language of Special Relativity (SR).

 However, energy conservation and Planck's equation predict a time dilation depending on the position in space also for non-gravitational  potentials. Thus, there should be also an influence of the non gravitational potentials on spacetime. This is achieved by the RND model. We expect that this theory will be able to describe properly dynamics for the microscopic region (which differ significantly from the classical one) because of extremely high acceleration and non gravitational forces playing the main role there. The observed precession and non-linearity of the dynamics equation of a binary predicts \cite{F16} chaotic motion of an electron in a hydrogen-like atom.

\acknowledgments { We wish to thank the referee for his valuable suggestions that helped to improve the paper considerably. Many thanks to Dr. Richard Manchester of the Australia Telescope Facility for providing information about the double pulsar. Finally,} we wish to acknowledge Dr. Rosemary Mardling of Monash University for suggesting to test our model for the  Hulse-Taylor pulsar's periastron advance.

\end{document}